\documentclass[11pt]{article}

\usepackage{amssymb}
\usepackage{amsmath}
\usepackage{mathrsfs}
\usepackage[numbers,sort&compress]{natbib}
\usepackage{graphicx}
\usepackage{setspace}
\usepackage{subfigure}
\usepackage{hyperref}
\usepackage{xspace,soul}

\newcommand{\pkg}[1]{{\texttt{#1}}}
\newcommand{\R}{\textsf{R}\xspace}

\begin{document}
\begin{center}
\textbf{Greater data science at baccalaureate institutions} \\ Amelia McNamara, Nicholas J Horton, Benjamin S Baumer \\ \today
\end{center}

Donoho's paper is a spirited call to action for statisticians, who he points out are losing ground in the field of data science by refusing to accept that data science is its own domain. (Or, at least, a domain that is becoming distinctly defined.) He calls on writings by John Tukey, Bill Cleveland, and Leo Breiman, among others, to remind us that statisticians have been dealing with data science for years, and encourages acceptance of the direction of the field while also ensuring that statistics is tightly integrated. 

As faculty at baccalaureate institutions (where the growth of undergraduate statistics programs has been dramatic~\citep{ASA2015}), we are keen to ensure statistics has a place in data science and data science education. In his paper, Donoho is primarily focused on graduate education. At our undergraduate institutions, we are considering many of the same questions.


 


We enthusiastically concur with Donoho's description of a ``Greater Data Science'' comprised of 
\begin{enumerate}
\item Data Gathering, Preparation, and Exploration 
\item Data Representation and Transformation 
\item Computing with Data
\item Data Modeling
\item Data Visualization and Presentation
\item Science about Data Science
\end{enumerate}
and aim to have our students develop all these key capacities in our courses and major programs.

In considering our curriculum development, we have been guided by the 2014 American Statistical Association (ASA)'s \emph{Curriculum Guidelines for Undergraduate Programs in Statistical Science}~\citep{asa-guidelines} and the 2016 \emph{Guidelines for Assessment and Instruction
in Statistics Education (GAISE) College Report}~\citep{CarEve2016}. Both documents highlight the need for students to work with real problems, messy data, and complex models.

Even more recently, a working group (including Baumer) developed the \emph{Curriculum Guidelines for Undergraduate Programs in Data Science}, which have now been endorsed by the ASA~\citep{pcmi2016guidelines}. This forward-thinking document addresses one of Donoho's primary concerns with data science education---that it may end up being a piecemeal collection of extant courses, with little ``long-term direction.'' While~\citep{pcmi2016guidelines} does provide guidance to institutions working with existing courses, it also outlines a model curriculum with a number of new and reformulated courses. 

\section*{Data science developments at our institutions}

Both the Smith College major in statistical and data sciences and the Amherst College major in statistics have been explicitly structured to introduce, extend, and integrate work in all six of the areas of Greater Data Science. Real problems have been interwoven into our courses at multiple levels. This has required extensive revision of existing courses along with the creation of a number of new and courses with complementary learning outcomes. 

At both Smith College and Amherst College, the introductory course touches on all six GDS elements, with an increased emphasis on visualization and modeling~\citep{RJ-2017-024, BauCet2014}. In subsequent courses like \emph{Multiple Regression} or \emph{Intermediate Statistics}, students explore, prepare, clean, transform, and visualize data. In the \emph{Communicating with Data}, \emph{Visual Analytics}, and \emph{Multivariate Data Analysis} courses, students learn principles of data visualization and presentation of data. Modeling is reinforced in \emph{Multiple Regression} and \emph{Machine Learning}. Capstone courses help to integrate prior course work with project-based learning while further refining computing and communication skills.

Existing Amherst College theory courses such as \emph{Probability and Theoretical Statistics} have been restructured to integrate computing as an explicit learning outcome (e.g., how to write a function, how to perform simulations, how to undertake empirical problem solving to complement analytic results, and how to collaborate in groups using GitHub). 

At Smith College, \emph{Introduction to Data Science}, \emph{Communicating with Data}, \emph{Visual Analytics}, and \emph{Machine Learning} are all new offerings guided by our understanding of data science as its own discipline. 

We would like to draw particular attention to \emph{Introduction to Data Science}, a successor to the course described in~\citep{Bau2015} that is offered at both institutions. Donoho makes reference to this course, which teaches data visualization, data wrangling, ethics, SQL, and communication, using a new textbook~\citep{BauHor2016}. The course is tied together by \emph{liberal arts modules}, where professor from other disciplines outline a question relevant to their discipline, and the students seek to address it using their new-found data skills. 

As Donoho reminds us, some academic statisticians have long been guilty of eschewing data analysis. But even some programs in data science focus more on tools and skills rather than developing the capacity to solve real problems. We believe our positions at liberal arts colleges give us a particular ability to reach across disciplines, connecting to data in the sciences, social sciences, and the humanities. The integration of liberal arts modules in \emph{Introduction to Data Science} can be used as a model for similar courses.

Another learning outcome in all of our courses is to produce students who learn how to learn. As with many disciplines, data science is evolving quickly. The tools we teach our students today may not be relevant in five years. In fact, several of the R packages referenced by Donoho (\pkg{reshape2} and \pkg{plyr}) have now been supplanted by others (\pkg{tidyr} and \pkg{dplyr})~\citep{dplyr, tidy-data}. As instructors, we do our best to stay on top of the current computational trends to provide our students with the most contemporary methods, which requires us to continually modify our curriculum. However, the focus is on generalized problem-solving that can be applied using different tools in different settings.  

Ethical precepts are an important part of any data science program. Donoho alludes to this with his detailed coverage of the University of California--Berkeley Master's program, which includes a course now titled ``Behind the Data: Humans and Values'' (formerly ``Legal, Policy, and Ethical Considerations for Data Scientists'')~\citep{BerkeleyMasters2017}. At Amherst ethics is now included as a learning outcome in the \emph{Intermediate Statistics} course with subsequent extension and reinforcement in elective and capstone courses. Ethics is also a component of the \emph{Introduction to Data Science} courses. Students consider questions like those posed in~\citep{boyCra2012}: what are the ethical implications of data science products? Who has access to data science, and who does not? What are our ethical obligations to our clients, ourselves, and our subjects? These higher-level questions make up a key part of the capstone courses.

At all levels, our courses emphasize best practices of statistical computing and reproducible research. These efforts build upon scholarly work that goes back at least to Don Knuth's literate programming~\cite{knut:1992} and Donoho's previous work on reproducibility~\cite{buckheit1995wavelab}. Baumer and McNamara are former faculty fellows on Project TIER: Teaching Integrity in Empirical Research~\cite{ball2012teaching}, which aims to spread good computing and data practices to the social sciences.  We are now seeing evidence of adoptions at our institutions, and others, where faculty members in economics, psychology, and environmental science and policy integrate reproducible research into their coursework, further strengthening our pool of data-capable students.


\section*{Data science scholarship}

Beyond our interest in the pedagogy of data science, we are also researchers. 
However, this is an area that is also undergoing development. Since it is an emerging field, institutions must determine how to judge new types of scholarly production. Like many problems of data science, this is something that applied statisticians have been wrestling with for decades. However, not all data science work is precisely applied statistics (thus, the new degree programs and scholarship).



Much like Donoho's notion of Science about Data Science, Jeff Leek has been proposing the idea of Data Science as a Science~\citep{Lee2016}. While Donoho's examples focus on meta-analysis, Leek's conception includes hands-on research. Calling on examples like Cleveland's study of graphical perception~\citep{cleveland1985graphical}, Leek advocates for data scientists experimenting to learn how software syntax impacts learning, and how practitioners are actually working (like~\citep{silberzahn2017many}).  

As a case study of scholarly production in data science, consider Hadley Wickham's many contributions. Wickham's work often centers on a profoundly useful \R package. However, each piece of software fits into a higher-level framework of intellectually-weighty ideas. The ideas behind \pkg{ggplot2} were articulated in a book on implementing Wilkinson's Grammar of Graphics~\cite{Wil2005, ggplot2}.  In addition to \pkg{tidyr}, Wickham wrote a article in the \emph{Journal of Statistical Software} on the concept of tidy data, which transcends the language it is implemented in~\cite{tidy-data}. Although these works are highly-cited, they do not fit cleanly into the traditional fields of statistics (having nothing to do with modeling, estimation, or inference) nor computer science (software engineering?).  We submit that these are early, influential works of scholarship in data science. 

Another set of exemplary papers can be found in a recently-published collection of articles---curated by Jenny Bryan and Hadley Wickham---entitled \emph{Practical Data Science for Stats} (to which the authors all contributed)~\citep{wickham2017peerj}. These articles discuss meta-data science topics like how to package reproducible analytical work~\citep{marwick2017packaging}, how to organize data in a spreadsheet~\citep{broman2017organization}, how to share data for collaboration~\citep{ellis2017sharing}, and how to implement a version control system~\citep{bryan2017git}. Our contributions discussed surviving as an isolated data scientist~\citep{baumer2017lessons}, and wrangling categorical data~\citep{mcnamara5wrangling}. 

The collection also contains an article on evaluating scholarly work in data science, focusing particularly on data science faculty in traditional statistics and biostatistics departments~\citep{waller2017}. Can these exemplary scholarly contributions in data science be neatly categorized into statistics or computer science research? If not, this further strengthens the notion that data science exists as a field of research unto itself.



\section*{Situating greater data science}

This brings us to our final question. If Donoho's vision of `Greater Data Science' takes hold, one wonders whether the current academic departmental alignments will (or should) continue. Of the authors, one is situated within a Department of Mathematics and Statistics (Horton), while the other two are appointed in a Program in Statistical and Data Sciences. Which approach is most fruitful? 

Clearly, there are many other academic areas that use data and data science methods. As we've discussed, our colleagues across the disciplines are embracing it. However, if data science is its own discipline, it cannot be solely situated within data-generating departments. Its unique teaching and scholarship indicate it may need to become a separate entity.


\bibliographystyle{apalike}
\bibliography{DonohoBib.bib}

\end{document}